\documentclass{article}
\usepackage{spconf,amsmath,graphicx}
\usepackage{color}
\usepackage{booktabs}  %  引入三线表宏包
\usepackage{amsfonts}
\usepackage{algorithm}
\usepackage{algorithmic}
\usepackage{multirow}
\usepackage{soul}
\usepackage{pgfplots}
\usepackage{tikz}
\usepackage{subfigure}

\DeclareMathOperator*{\argmax}{arg\,max}
\newcommand{\etal}{\textit{et al}.}

\newcommand{\eg}{\textit{e}.\textit{g}.}
\title{AUC Optimization for Robust Small-footprint Keyword Spotting with Limited Training Data}
\name{Menglong Xu, Shengqiang Li, Chengdong Liang, Xiao-Lei Zhang \thanks{The email of the first author is mlxu@mail.nwpu.edu.cn.}}

\address{CIAIC, School of Marine Science and Technology, Northwestern Polytechnical University, China}

\begin{document}
%\ninept

\maketitle

\begin{abstract}
  Deep neural networks provide effective solutions to small-footprint keyword spotting (KWS).
  However, if training data is limited, it remains challenging to achieve robust and highly accurate KWS in real-world scenarios where unseen sounds that are out of the training data are frequently encountered.
  Most conventional methods aim to maximize the classification accuracy on the training set, without taking the unseen sounds into account.
  To enhance the robustness of the deep neural networks based KWS, in this paper, we introduce a new loss function, named the maximization of the \textit{area under the receiver-operating-characteristic curve} (AUC).
  The proposed method not only maximizes the classification accuracy of keywords on the closed training set, but also maximizes the AUC score for optimizing the performance of non-keyword segments detection.
  Experimental results on the Google Speech Commands dataset v1 and v2 show that our method achieves new state-of-the-art performance in terms of most evaluation metrics.
\end{abstract}

\section{Introduction}
\label{sec:intro}
  Keyword spotting (KWS), also known as spoken term detection (STD), is the task of detecting some predefined keywords from a stream of utterances. It is usually used as an intelligent agent in mobile phones or smart devices.
  Recently, deep neural network (DNN) based KWS has led to significant performance improvement over conventional methods.
  Deep KWS \cite{chen2014small} first considers keyword spotting as an audio classification problem.
  It trains a DNN model to predict the posteriors of predefined keywords, in which each neuron in the softmax output layer of the DNN model corresponds to a keyword, with an additional ``filler" neuron representing all other non-keyword segments.
  This classification-based method achieves significant improvement over the keyword/filter hidden Markov models.
  Later on, a number of classification-based methods \cite{arik2017convolutional,tang2018deep,shan2018attention,choi2019temporal,bai2019time,xu2020depthwise,yang2020multi} were explored to miniaturize the memory footprint.

  However, because the softmax cross entropy loss focuses on maximizing the classification accuracy of the training data, the aforementioned models require a large number of training samples to achieve robust performance against various non-keyword segments in the test stage \cite{chen2014small,arik2017convolutional,shan2018attention}.
  Because collecting as many types of non-keyword segments as possible for the model training is expensive and sometimes unavailable,
  the classification-based models \cite{tang2018deep,choi2019temporal,bai2019time,xu2020depthwise,yang2020multi} perform particularly poor in practice.
  Moreover, using a single ``filler" neuron to represent all non-keyword segments does not reflect the diversity between these sounds, which will further degrade the performance.
  %\cite{huh2021metric}.

  Recently, several works \cite{sacchi2019open,yuan2019verifying,zhang2020deep,huh2021metric,vygon2021learning} introduced metric learning into KWS.
  Metric learning adopts a ranking loss to learn the relative distance between samples. It aims to enlarge the inter-class variance and reduce the intra-class variance in an embedded space of data.
  However, it will result in a significant performance drop if we directly apply metric learning to KWS without taking the prior knowledge that the target keywords are predefined and fixed into consideration.
  %However, \textcolor[rgb]{0.00,0.50,1.00}{if we} \textcolor[rgb]{1.00,0.00,0.00}{\st{these works}} directly applied metric learning to KWS without taking
%  %However, directly applying metric learning to KWS ignores
%  the prior knowledge that the target keywords are predefined and fixed into consideration, \textcolor[rgb]{0.00,0.50,1.00}{it will} \textcolor[rgb]{1.00,0.00,0.00}{\st{which}} result\textcolor[rgb]{1.00,0.00,0.00}{\st{s}} in a significant performance drop \cite{huh2021metric}.
  To address the problem,
  Huh \etal \ \cite{huh2021metric} proposed an angular prototypical network with fixed target classes (AP-FC) to enhance the robustness against non-keyword segments. However, they have to use an additional support vector machine (SVM) to make the final decision.
  In \cite{vygon2021learning}, Vygon \etal \ combined a triplet loss-based embedding extractor with a K-Nearest Neighbor (kNN) classifier, which gets higher accuracy than the cross entropy loss based methods. Their method exceedingly increases the number of parameters and computational complexity of the KWS model.

  Motivated by some works on the open-set recognition problem \cite{bendale2016towards,devries2018learning}, in this paper, we propose a new loss function, named the maximization of the \textit{area under the receiver-operating-characteristic curve} (AUC), and a confidence based decision method, which leads to a robust, small-footprint, and high accuracy KWS model.
  Specifically, the proposed multi-class AUC loss maximizes the classification accuracy of predefined keywords, and the detection AUC of non-keyword segments simultaneously.
  We compared the proposed multi-class AUC loss with softmax cross entropy loss \cite{tang2018deep}, prototypical loss \cite{huh2021metric}, AP-FC loss \cite{huh2021metric}, and triplet loss \cite{vygon2021learning} on the Google Speech Commands dataset v1 \cite{warden2017speech} and v2 \cite{warden2018speech}.
  Experimental results demonstrate that our methods outperform the comparison methods in most evaluation metrics.
  The main contributions of this paper are summarized as follows:
  \begin{itemize}
    \item To our knowledge, we reformulate the low resource keyword spotting task as an open-set recognition problem for the first time.
    \item We propose a novel multi-class AUC loss. It outperforms the four representative referenced methods in most evaluation metrics.
    \item We propose a new confidence-based decision method. It helps the proposed method achieve the state-of-the-art performance without using a complex back-end classifier.
  \end{itemize}

  The remainder of the paper is organized as follows.
  Section \ref{sec:algorithm} introduces the proposed AUC loss.
  Section \ref{sec:experiments} and \ref{sec:results} present the experimental setup and results respectively.
  Section \ref{sec:conclusions} concludes the paper.

  \section{Background}
  The original AUC optimization is designed for binary-class classification only. Therefore, before describing the proposed multi-class AUC loss function, we first take a look at the existing binary AUC optimization.

Given a binary-class dataset $\mathbf{X}=\left\{\left(\mathbf{x}_{n}, y_{n}\right)\right\}_{n=1}^{N}$ where $y_n\in\{0,1\}$, and a binary-class neural network $f_{\theta}(\cdot)$ with $\theta$ being the parameter of the network, we define two new subsets: $\mathbf{S}^{+}=\left\{f_{\theta}(\mathbf{x}_n),\ \ \forall \mathbf{x}_n \in \mathbf{X} \mid y_{n}=1\right\}$ which is a set of neural network scores for the samples with $y_n = 1$, and $\mathbf{S}^{-}=\left\{f_{\theta}(\mathbf{x}_n), \ \ \forall \mathbf{x}_n \in \mathbf{X} \mid y_{n}=0\right\}$ which represents a set of neural network scores for the samples with $y_n = 0$.
   Cardinalities of these two subsets are $N^{+}$ and $N^{-}$ respectively.
   As described in \cite{Fan2019vad}, for the finite set of samples $\mathbf{X}$, the approximate estimate of the AUC metric is:
   \begin{equation}\label{bauc}
     \text{AUC}=\frac{1}{N^{+} N^{-}} \sum_{i=1}^{N^{+}} \sum_{j=1}^{N^{-}} \mathbb{I} \left(s_{i}^{+}>s_{j}^{-}\right)
   \end{equation}
   where $\mathbb{I}(\cdot)$ is an indicator function that returns 1 if the statement is true, and 0 otherwise, and $s_{i}^{+}$ and $s_{j}^{-}$ are the elements of $\mathbf{S}^{+}$ and $\mathbf{S}^{-}$ respectively.
   As \cite{bai2020partial} did, we relax (\ref{bauc}) by replacing the indicator function by a modified hinge loss function:
   \begin{equation}\label{hinge}
     \ell_{\text{hinge}}^{\prime}(z)=\max (0, \delta-z)^{2}
   \end{equation}
   where $z=s_{i}^{+}-s_{j}^{-}$, and $\delta > 0$  is a tunable hyperparameter controlling the distance margin between $s_{i}^{+}$ and $s_{j}^{-}$.
   Substituting (\ref{hinge}) into (\ref{bauc}) transforms the maximization problem of (\ref{bauc}) into the following minimization problem:
   \begin{equation}\label{bloss}
    \ell=\frac{1}{N^{+} N^{-}} \sum_{i=1}^{N^{+}} \sum_{j=1}^{N^{-}} \max \left(0, \delta-\left(s_{i}^{+}-s_{j}^{-}\right)\right)
   \end{equation}
   which can be easily backpropagated throughout the network in a standard procedure.

\section{Algorithm description}
\label{sec:algorithm}
  %In this section, we first describe the commonly used binary and multiclass AUC metrics and then present our modification.

  \subsection{Problem formulation}
  \label{sec:formulation}
  In this paper, we decompose the KWS task into a non-keyword segments detection subtask and a closed-set classification subtask.
  Specifically, for a given input sample, we first determine whether it belongs to a predefined keyword set. If so, then we decide which keyword it is. Note that the two subtasks are performed simultaneously in our proposed method.

  To formalize the task, suppose there is a dataset $\mathbf{X}=\left\{\left(\mathbf{x}_{n}, y_{n}\right)\right\}_{n=1}^{N}$ where $\mathbf{x}_{n} \in \mathbb{R}^{D}$ is a high-dimensional acoustic feature of the $n$-th sample,
  and $y_{n} \in \{0,1,2,\dots,C\}$ is the ground-truth label of $\mathbf{x}_{n}$. Note that, without loss of generality, we always assume that there are $C+1$ categories with class $0$ representing non-keyword segments, and the other classes $1, 2, \dots, C$ representing $C$ keywords respectively.

 We aim to train a neural network $f_{\theta}(\cdot):\mathbb{R}^{D}\to\mathbb{R}^{C}$ where $\theta$ is the parameter of the network. It maps the $D$-dimensional input acoustic feature to a $C$-dimensional vector. Each dimension of the vector represents the confidence score of its corresponding keyword.
  In the test stage, we use $f_{\theta}(\cdot)$ to conduct KWS by the following criterion:
  \begin{equation}\label{criterion}
    \hat{y}_n=\left\{
     \begin{array}{ll}
     \argmax_{c} \left([p_{n,c}]_{c=1}^{C}\right),
     & \text{if } \max_{c} \left([p_{n,c}]_{c=1}^{C}\right) \geq \eta \\
     0, & \text{otherwise}
    \end{array}\right.
  \end{equation}
  where $[p_{n,c}]_{c=1}^{C}=[p_{n,1},p_{n,2},\dots,p_{n,C}]^T$ is the output scores of the neural network $f_{\theta}(\mathbf{x}_n)$, and $\eta$ is a decision threshold. For simplicity, we denote $\mathbf{p}_n = [p_{n,c}]_{c=1}^{C}$ in the remaining of the paper.

  \subsection{The proposed multi-class AUC optimization}

   Several studies have extended the binary AUC optimization to multi-class problems \eg \  \cite{bai2020partial,gimeno2021generalising}.
   In this work, we propose a new extension suitable for most multi-class classification tasks and computationally straightforward. The key idea of this extension is to modify the two subsets $\mathbf{S}^{+}$ and $\mathbf{S}^{-}$ in the binary AUC optimization to new forms that satisfy the multi-class AUC optimization problem.

   Specifically, for the general KWS problem with more than one keyword, we define the subset of positive examples as
   $$\mathbf{S}^{+}=\left\{p_{n,y_n}, \ \ \forall \mathbf{x}_n \in \mathbf{X} \mid y_{n} \in \{1,2,\dots,C\} \right\}$$
   and the subset of negative samples $\mathbf{S}^{-}=\ \mathbf{S}_{1}^{-} \cup\ \mathbf{S}_{2}^{-}$ with
   $$
    \mathbf{S}_{1}^{-}=\left\{\max_{c,c\neq y_n} \left([p_{n,c}]_{c=1}^{C}\right), \ \ \forall \mathbf{x}_n \in \mathbf{X} \mid y_{n} \in \{1,2,\dots,C\}\right\}
   $$
   where $p_{n,y_n}$ is the score at the $y_n$-th position of the vector $\mathbf{p}_n$, $\max_{c,c\neq y_n} \left([p_{n,c}]_{c=1}^{C}\right)$
   is the maximum value of $\mathbf{p}_n$ after removing the score at the $y_n$-th position of $\mathbf{p}_n$, and
   $$
    \mathbf{S}_{2}^{-}=\left\{\max_{c} \left([p_{n,c}]_{c=1}^{C}\right), \ \ \forall \mathbf{x}_n \in \mathbf{X} \mid y_{n}=0\right\}
   $$
   represents the set of the output scores of the neural network for the non-keyword segments in $\mathbf{X}$.

   Algorithm \ref{mauc} presents the proposed multi-class AUC loss in detail.
   \begin{algorithm}[t]

    \caption{Multi-class AUC loss for KWS}
    \label{mauc}
    \begin{algorithmic}[1] %这个1 表示每一行都显示数字
     \REQUIRE ~~\\ %算法的输入参数：Input
      a batch of acoustic features, $\mathbf{x}$; \\
      the corresponding labels, $\mathbf{y}$; \\
      the number of samples in the mini-batch, $N$; \\
      predefined hyperparameter, $\delta$; \\
     \ENSURE ~~\\ %算法的输出：Output
      loss $\ell$ on the current mini-batch;
     \vspace{9pt}
     \STATE $N^{+}\leftarrow \sum_{n=1}^{N}\mathbb{I}(y_n\neq0)$;
     \STATE $N^{-}\leftarrow N$;
     \STATE Init the positive subset $S^+$ which contains $N^+$ samples and the negative subset $S^-$ which contains $N^-$ samples;
     \STATE $\mathbf{p}\leftarrow f_{\theta}(\mathbf{x})$;
     \FOR{each $y_{n} \in \mathbf{y}$}
      \IF{$y_n \neq 0$}
       \STATE add $p_{n,y_n}$ to subset $S^+$;
       \STATE add the largest element of $\mathbf{p}$ except $p_{n,y_n}$ to subset $S^-$;
      \ELSE
       \STATE add the largest element of $\mathbf{p}$ to subset $S^-$;
      \ENDIF
     \ENDFOR
     \STATE $\ell=\frac{1}{N^{+} N^{-}} \sum_{i=1}^{N^{+}} \sum_{j=1}^{N^{-}} \max \left[0, \delta-\left(s_{i}^{+}-s_{j}^{-}\right)\right]$;
     \vspace{2pt}
     \RETURN $\ell$; %算法的返回值
    \end{algorithmic}
   \end{algorithm}

   \subsection{Confidence based decision for the multi-class AUC loss}

   In the test stage, the decision threshold $\eta$ is calculated on a validation set by:
   \begin{equation}\label{eta}
     \eta=- \delta + \frac{1}{\sum_{n=1}^{N^{\prime}}\mathbb{I}(y_n\neq0)} \sum_{n=1}^{N^{\prime}} {\mathbb{I}(y_n\neq0)\ p_{n,y_n}}
   \end{equation}
   where $N^{\prime}$ is the size of the validation set.

  \subsection{Connection to other loss functions}

  This subsection presents the connection of the proposed multi-class AUC to other loss functions.

   \subsubsection{Connection to multi-class hinge loss}

   Under the same supposition in Section \ref{sec:formulation}, the multi-class classification hinge loss is presented as:
   \begin{equation}
    \label{hinge_loss}
    \ell_{\text{hinge}}=\frac{1}{NC} \sum_{n=1}^{N} \sum_{\substack{\{c|c=0, \ldots,\\C, \ \ c\neq y_n\}}} \max (0, \delta-p_{n,y_{n}}+p_{n,c})
   \end{equation}

   The connection between the proposed multi-class AUC loss and the multi-class hinge loss is as follows.
   The  multi-class AUC loss calculates the loss on the whole training set. It essentially learns a rank of the training samples without resorting to a classification-based loss explicitly.
   In contrast, the multi-class hinge loss calculates the optimization objective on each sample respectively and then averages them on the entire dataset. It needs to assign all non-keyword segments to a single class.

   \subsubsection{Connection to AP-FC loss}

The AP-FC loss first arranges the keywords in a predefined order. Then, for each mini-batch, it selects one sample from each keyword, followed by $N_0$ non-keywords. Note that the first $C$ samples should be arranged in the predefined order of the keywords.

   According to \cite{huh2021metric}, we rewrite the AP-FC loss as:
   \begin{eqnarray}
     \ell_{\text{AP-FC}} &=& -\frac{1}{C} \sum_{c=1}^{C} \log \frac{e^{\mathbf{S}_{c,c}}}{\sum_{n=1}^{C} e^{\mathbf{S}_{n, c}} +\sum_{k=1}^{N_0} e^{\mathbf{S}_{C+k, c}}} \notag \\
      &=& -\frac{1}{C} \sum_{c=1}^{C} \log (\text{softmax}(\mathbf{S}^T))_{c,c}
   \end{eqnarray}
   with
   \begin{equation}
    \mathbf{S}_{n,c}=w \cos \left(\mathbf{e}_{n}, \mathbf{W}_{c}\right)+b, \ c \in \{1,2,\dots,C\}
   \end{equation}
   where $\mathbf{e}_{n}$ is the extracted feature of the $n$-th sample by the neural network, $\mathbf{W}_c$ is the learnable class center of the $c$-th keyword, and $w, b$ are learnable parameters with $w>0$.

   The proposed AUC loss and AF-FC loss are similar in that they do not assign widely distributed non-keyword segments to a single ``filler" class.
   However, the implementation of the AP-FC loss has a strict constraint on the samples in each mini-batch.
   %This inefficient data sampling method may result in a
   Moreover, the AP-FC loss-based model still needs an SVM back-end to make the final decision.

   \subsubsection{Connection to other multi-class AUC loss}

   The multi-class AUC optimization in \cite{gimeno2021generalising} is a natural extension of the binary AUC optimization.
   Gimeno \etal\ extended the binary AUC optimization to the multi-class problem by the one-versus-one and one-versus-rest frameworks.
  The one-versus-one multi-class AUC loss is obtained by averaging the pairwise binary AUC losses.
   The one-versus-rest multi-class AUC loss decomposes the multi-class classification task to $C$ binary tasks. For the $c$-th task, the $c$-th class is viewed as a positive class, and all other classes are merged into a negative class.
   However, the above two methods cannot be directly used for our open-set optimization problem, since that they need to assign non-keyword segments to a ``filler" class.
   In addition, it is obvious that our proposed AUC loss is more computationally efficient than the above two methods.

 \section{Experimental setup}
 \label{sec:experiments}

  \subsection{Data preparation}
   In our experiments, two popular keyword spotting datasets, Google Speech Commands version 1 (GSC v1) \cite{warden2017speech} and version 2 (GSC v2) \cite{warden2018speech} are used for evaluation.
   The dataset GSC v1 consists of 65K one-second-long recordings of 30 words from thousands of different speakers.
   GSC V2 is an augmented version of GSC v1, which contains 105K utterances of 35 words.
   In addition, both datasets contain several minute-long background noise files.
   The sampling rates of all signals are 16 kHz in the two datasets.

 %Both GSC v1 and GSC v2 include a ``validation\_list" file, which contain a list of files that are expected to be used for validating.
% The ``testing\_list" file contains the names of audio clips that should only be used for testing.
 Both GSC v1 and GSC v2 include a ``validation\_list" file and a ``testing\_list" file.
 We use audio files in the ``validation\_list" and ``testing\_list"  as validation and testing data, and the other audio files as training data.
 Following previous works, we apply random time-shift and noise injection to training data.
 Specifically, we first perform a random time-shift of $t$ milliseconds to each sample, where $t\sim U[-100,100]$.
 We then add background noise to each sample with a probability of 0.8, where the noise is chosen randomly from the background noises.
 Note that the random time-shift and noise injection are performed on the fly at each training step.
 Finally, 40-dimensional Mel-frequency Cepstrum Coefficient (MFCC) features are extracted and stacked over the time-axis with a window length of 25ms and a stride of 10ms.
 %In this section, we first describe the datasets and experimental settings.
 %%and evaluation metrices.
 %We then report and analyze the experimental results.

 \subsection{Backbone network}

\begin{figure}[t]
     \centering
     \includegraphics[scale=0.5]{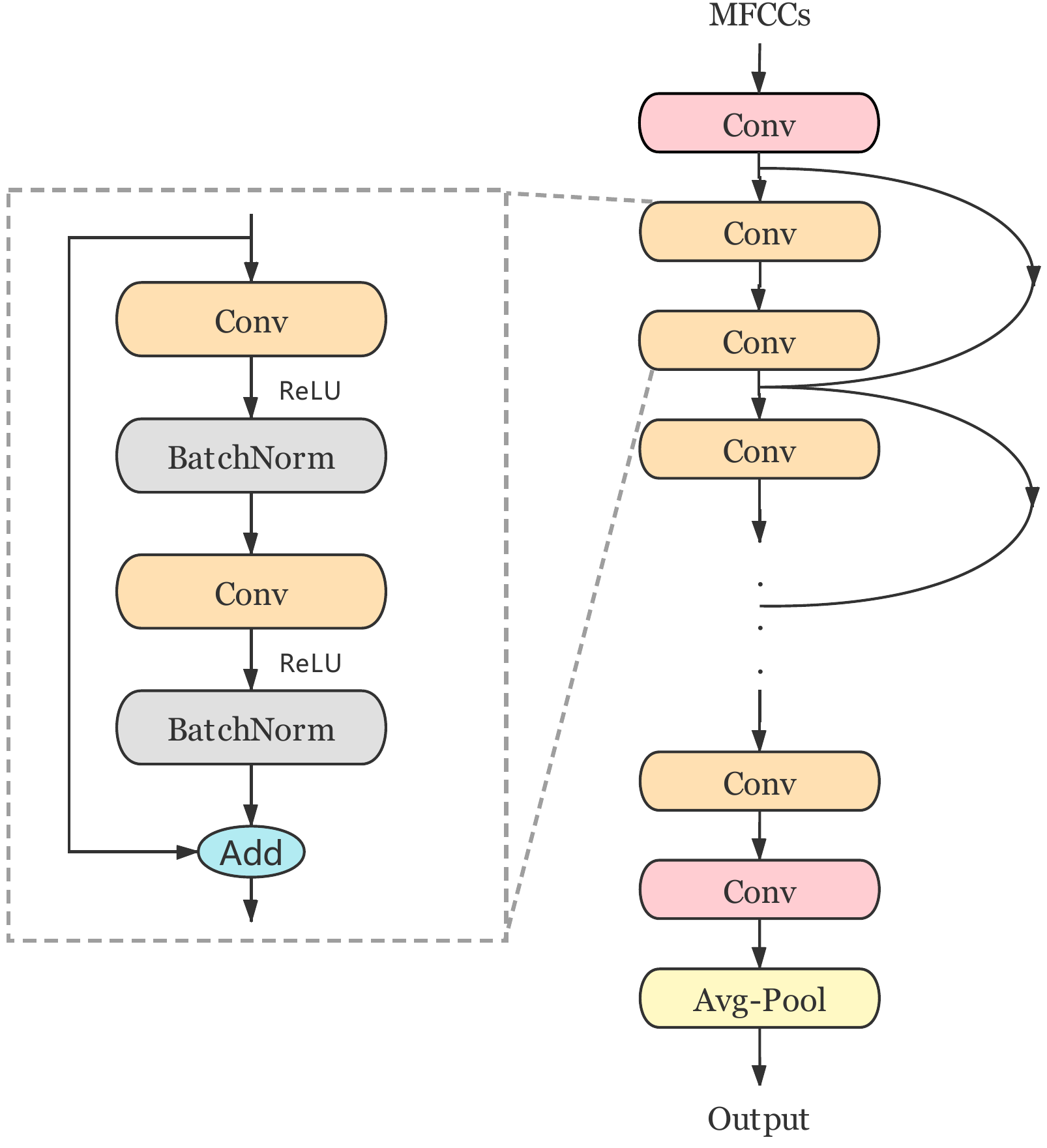}
     \caption{The architecture of the backbone network, with a magnified residual block.}
     \label{fig:model}
   \end{figure}

   We use \texttt{res15} \cite{tang2018deep} as the backbone network.
   As shown in Figure \ref{fig:model}, it starts with a bias-free convolution layer (Conv) with weight $\textbf{W}\in\mathbb{R}^{m\times r\times n}$, where $m$ and $r$ are the height and width of the convolution kernel respectively, and $n$ is the number of the output channels.
   Then, it takes the output of the first convolution layer as the input of a chain of residual blocks (Res), followed by a separate non-residual convolution layer.
   Finally, the output of the network is obtained by an average-pooling layer (Avg-pool).
   Additionally, a $(d_{w},d_{h})$ convolution dilation is used to increase the receptive field of the network, and a batch normalization layer (BatchNorm) is added after each convolution layer to help train the deep network.
   The details of the backbone network are listed in Table \ref{table:res15}.
   \begin{table}[t]
      \caption{Parameter setting of \texttt{res15}, along with the number of parameters and multiplies.}
      \hspace{8pt}
      \label{table:res15}
      \centering
      \vspace{6pt}
      \scalebox{0.98}{
      \begin{tabular}{c | c c c c c | c c}
        \hline
         &
        \multicolumn{1}{c}{$m$} &
        \multicolumn{1}{c}{$r$} &
        \multicolumn{1}{c}{$n$} &
        \multicolumn{1}{c}{$d_{w}$} &
        $d_{h}$ &
        \multicolumn{1}{c}{\#Par.} &
        \multicolumn{1}{c}{\#Mult.} \\
        \hline
        Conv & 3 & 3 & 45 & 1 & 1 & 405 & 1.52M
        \\
        Res$\times$6 & 3 & 3 & 45 &
        $2^{\lfloor\frac{i}{3}\rfloor}$ & $2^{\lfloor\frac{i}{3}\rfloor}$ & 219K & 824M
        \\
        Conv & 3 & 3 & 45 & 16 & 16 & 18.2K & 68.6M              \\
        BatchNorm & - & - & 45 & - & - & - & 169K              \\
        Avg-Pool & - & - & 45 & - & - & - & 45               \\
        \hline\hline
        Total  & - & - & - & - & - & 238K & 894M
        \\
        \hline
      \end{tabular}
      }
   \end{table}

    \begin{table*}[t]
 \centering
 \caption{Comparison results between the proposed multi-class AUC and four referenced methods. The subscript R indicates the \texttt{random sampler}, and F the \texttt{fixed proportion sampler}.}
 \vspace{12pt}
 \label{table:result}
 \begin{tabular}{l c c c c c c c c c}
 \toprule
 %\makecell{}
 \multirow{2}*{Loss} & \multirow{2}*{Back-end} & & \multicolumn{3}{c}{GSC v1} & & \multicolumn{3}{c}{GSC v2} \\
 \cline{4-6} \cline{8-10}
 & & & Total acc & Closed acc & F1 score & & Total acc & Closed acc & F1 score \\

 \hline
 Cross entropy \cite{tang2018deep} & - & & 89.96\% & 97.14\% & 0.8805 &  & 92.74\% & 97.46\% & 0.9068\\
 Prototypical \cite{huh2021metric} & - & & 87.89\% & 95.88\% & 0.8654 & & 93.32\% & 96.55\% & 0.9149 \\
 AP-FC \cite{huh2021metric} & SVM & & 91.59\% & 96.72\% & 0.8962 & &  93.77\% &  97.11\% & 0.9188  \\
 Triplet \cite{vygon2021learning} & kNN & & 92.09\% & 97.28\% & 0.9019 & &  94.01\% &  \textbf{97.78}\% &  0.9251 \\
 Multi-class AUC\footnotesize{R} & - & & 92.16\% & 97.01\% & 0.9031 & & \textbf{94.87}\% & 97.39\% & \textbf{0.9315} \\
 Multi-class AUC\footnotesize{F} & - & & \textbf{92.97}\% & \textbf{97.22}\% & \textbf{0.9115} & & 94.71\% & 97.50\% & 0.9312 \\
 \bottomrule
 \end{tabular}
\end{table*}

 \subsection{Tasks and evaluation metrics}
The tasks in previous works \cite{tang2018deep,choi2019temporal,xu2020depthwise,yang2020multi}
 focus on discriminating the 11 keywords (``yes", ``no", ``up", ``down", ``left", ``right", ``on", ``off", ``stop", ``go", ``silence") and a non-keyword ``unknown",
 %focus on distinguishing among 12 categories: ``yes", ``no", ``up", ``down", ``left", ``right", ``on", ``off", ``stop", ``go", ``silence" and ``unknown",
 where ``silence" denotes silence segments and ``unknown" represents all other words.
 In their settings, all unknown words used in the test set have been seen by the model in the training stage, which is not consistent with real-world KWS applications.

 To meet the real-world KWS applications, in our experiments, we consider the task in \cite{huh2021metric}, where ten unknown words (``zero", ``one", ``two", ``three", ``four", ``five", ``six", ``seven", ``eight", ``nine") are used for testing only.
 We evaluate our model by comparing the following metrics with other related works.
 \begin{itemize}
   \item \textbf{Total acc} is the classification accuracy on the test set that contains unseen unknown words, which is more likely to reflect the performance of the KWS model in the real world.
       Note that unseen unknown words represent the above ten unknown words that are used for testing only.
   \item \textbf{Closed acc} is the classification accuracy on the test set that does not contain unseen unknown words.
   %\item \textbf{Target Acc} is the classification accuracy on 11 keywords.
%   \item \textbf{Non-target Acc} is the fraction of unknown words that are correctly classified.
 %percentage of unknown words that are correctly classified to all unknown words.
   \item We also report the \textbf{F1 score} that is extended to multi-class one by ``macro" average on the test set
that contains unseen unknown words.
 \end{itemize}
 In addition, we plot the detection error tradeoff (DET) curve of the non-keyword segments detection subtask to evaluate the KWS models.

 \subsection{Data sampler}

Usually, the training samples in each mini-batch is randomly sampled from the whole training set, which results in the proportion of the keywords over non-keywords in each mini-batch vary greatly. We denote this sampling method as \texttt{random sampler}. However, the variable proportion will hinder the convergence of the model training of the proposed method. To overcome this problem, we use a \texttt{fixed proportion sampler}, which keeps the proportion of keywords and non-keywords consistent in each mini-batch.

 \subsection{Training details}

 Each model in our experiments is trained for 60 epochs, using the Adam optimizer \cite{kingma2014adam}.
 The initial learning rate is set to 0.001 and reduced to 0.0001 after 30 epochs.
 For the cross entropy loss, we use a mini-batch size of 128 and $L_2$ weight decay of $10^{-5}$.
 We use the same hyperparameters in \cite{huh2021metric} and \cite{vygon2021learning} for the prototypical loss, AP-FC loss and triplet loss.
 %For the Prototypical loss, AP-FC loss and Triplet loss, we use the same hyperparameters in \cite{huh2021metric} and \cite{vygon2021learning}.
 We use the validation set to select the best model among different epochs and evaluate the effect of the hyperparameter $\delta$.

 We evaluate the proposed multi-class AUC loss with the \texttt{fixed proportion sampler} and the \texttt{random sampler}.
 For the \texttt{fixed proportion sampler}, the number of keywords and non-keywords in each mini-batch is set to 32 and 64, respectively; for the \texttt{random sampler}, the mini-batch size is set to 128, which is the same as the other comparison methods.
 The hyperparameter $\delta$ is set to 0.3.
 Following the same training procedure, we evaluate all comparison methods for five independent times, and report the average performance.

\section{Results}
\label{sec:results}

 \subsection{Evaluation of the proposed methods}

Table \ref{table:result} lists the comparison result between the proposed methods and the four baselines. From the table, we see that both the two variants of the proposed multi-class AUC loss achieve significant improvement in terms of the Total acc and F1 score, and achieve a competitive result with the best referenced method in terms of the Closed acc.
 We take the result on GSC v1 as an example.
  Comparing to the cross entropy loss, the multi-class AUC loss with the \texttt{fixed proportion sampler} achieves 30.0\% and 25.9\% relative improvement in Total acc and F1 score respectively. It also achieves a slightly higher Closed acc than the cross entropy loss.
 Even when compared with the triplet loss with a complex kNN backend, the proposed method still achieves a relative improvement of 11.1\% in Total acc and 9.8\% in F1 score while maintaining a similar Closed acc.
 \begin{figure*}[tbp]
  \subfigure[Results on GSC v1.]{
   \hspace{-8pt}
   \begin{minipage}[b]{.5\linewidth}
    \centering
    \includegraphics[scale=0.36]{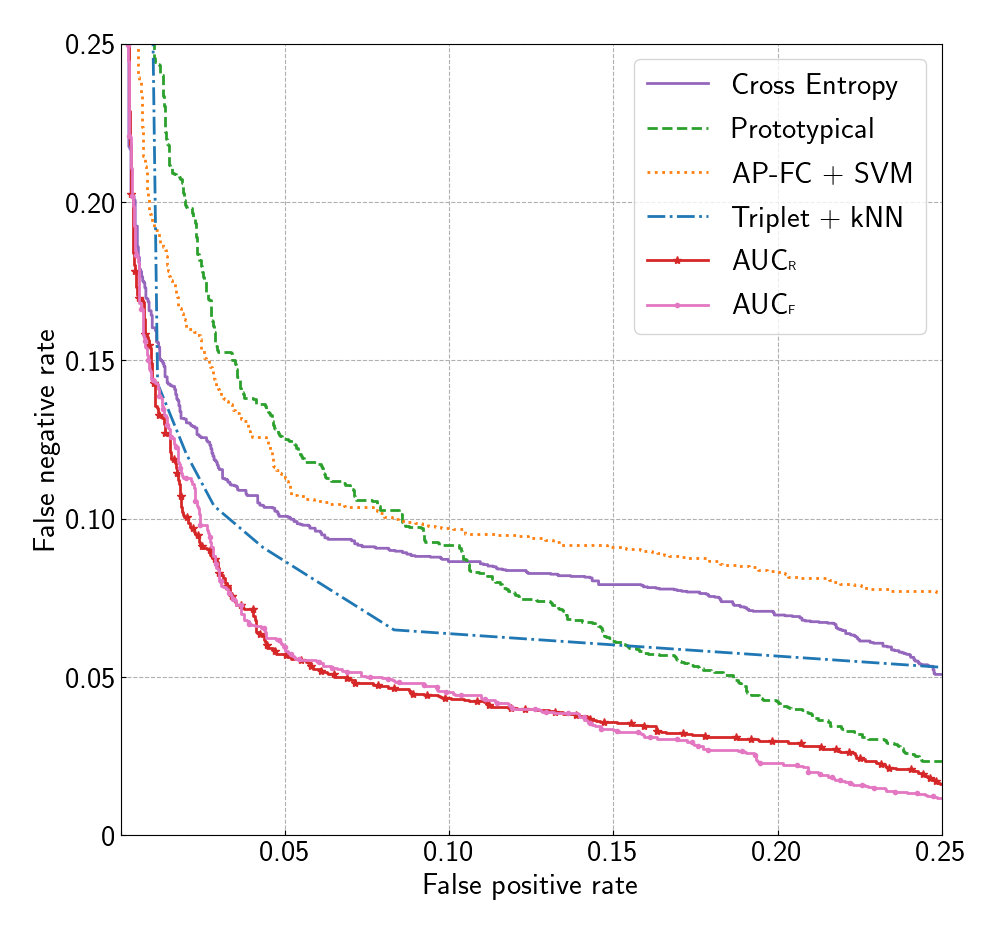}
   \end{minipage}
  }
  \subfigure[Results on GSC v2.]{
   \hspace{-16pt}
   \begin{minipage}[b]{.5\linewidth}
    \centering
    \includegraphics[scale=0.36]{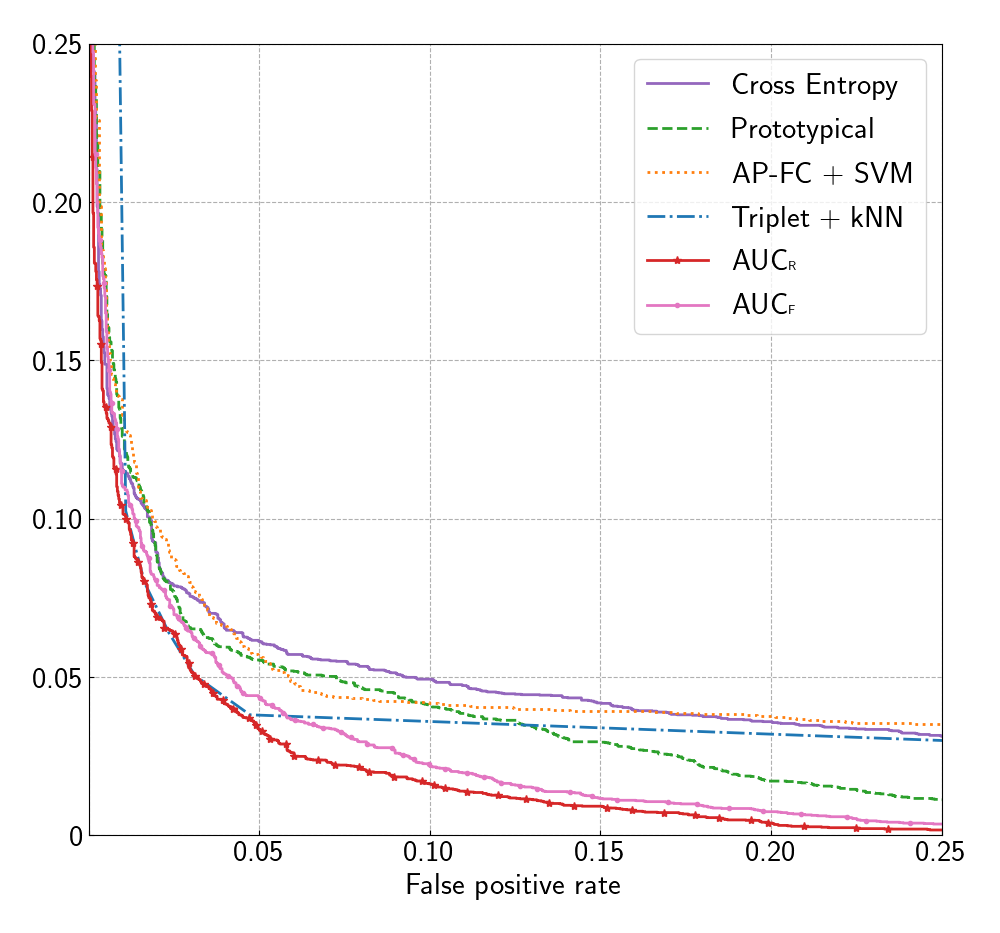}
   \end{minipage}
  }
  \vspace{-5pt}
  \caption{DET curves of the non-keyword segments detection subtask.}
  \label{fig:det}
 \end{figure*}

     \begin{table*}[t]
 \centering
 \caption{Effect of the hyperparameter $\delta$ on performance.}
  \label{table:delta}
 \vspace{10pt}
 \begin{tabular}{l c c c c c c c c c c}
 \toprule
   & & \multicolumn{7}{c}{AUC} & & \multirow{2}*{Cross Entropy} \\
 \cline{2-9}
 & sampler & 0.1 & 0.2 & 0.25 & 0.3 & 0.35 & 0.4 & 0.5 & &   \\

 \hline
 \multirow{2}*{Closed acc}  & R & 94.52\% & 96.29\% & 96.53\% & 96.85\% &  96.72\% &  96.49\% & 95.54\% & & \multirow{2}*{96.20\%}  \\
  & F & 94.04\% & 96.54\% & 96.71\% & 96.81\% &  96.44\% &  96.53\% &  96.20\% & & \\
 \multirow{2}*{F1 score} & R & 0.9426 & 0.9578 & 0.9581 & 0.9615 & 0.9577 & 0.9535 & 0.9429 & & \multirow{2}*{0.9513} \\
  & F & 0.9321 & 0.9599 & 0.9613 & 0.9613 & 0.9553 & 0.9546 & 0.9508 & &  \\
 \bottomrule
 \end{tabular}
\end{table*}

 To further investigate the effectiveness of the proposed method, we conduct a comparison on GSC v2 using the same settings as that on GSC v1.
 The experimental results again demonstrate the superiority of our method.
 In addition, the result on GSC v2 indicates that the training data of GSC v2 is responsible for the substantial improvement in all evaluation metrics, which is consistent with the experimental phenomenon in \cite{warden2018speech}.
 However, although both of the two variants of the proposed multi-class AUC loss achieve better results on GSC v2 than that on GSC v1, the improvement with \texttt{random sampler} is more evident than that with the \texttt{fixed proportion sampler}.
 This may be caused by that the training data of GSC v2 contains more non-keywords than the training data of GSC v1.

 From Table \ref{table:result} we also see that the AP-FC loss with a SVM back-end and the triplet loss with a kNN back-end outperform the prototypical loss and the cross entropy loss. It demonstrates that the metric learning-based methods still require a
 decision back-end to achieve satisfactory performance.
 In addition, we plot the DET curves of the non-keyword segments detection subtask in Figure \ref{fig:det}. From the figure, we see that these curves are consistent with the results presented in Table \ref{table:result}, and we see that the two variants of the proposed multi-class AUC loss outperform the referenced methods.
 %and achieve more competitive results than the cross entropy loss.

\subsection{Effect of the hyperparameter $\delta$ on performance}
 %\begin{figure}
%  \centering
%  % Requires \usepackage{graphicx}
%  \includegraphics[scale=0.456]{Fig/curve.png} \\
%  \caption{\textcolor[rgb]{1.00,0.00,0.00}{Effects of hyperparameters on performance}}\label{fig:curve}
% \end{figure}
 This subsection investigates the effect of the hyperparameter $\delta$ on performance.
 Becaue there are no unseen unknown words in the validation set, here we only use the Closed acc and F1 score as the evaluation metrics.
 For simplicity, we show the experimental results on GSC v1 only. Note that the experimental phenomenon on the other evaluation dataset are consistent with that on GSC v1.
Table \ref{table:delta} lists the result on GSC v1.
 From the table, one can see that the parameter $\delta$, which controls the margin of the AUC loss, plays an important role on the performance.
Both of the two variants of the multi-class AUC loss outperform the cross entropy baseline in the two evaluation metrics when $0.2\le\delta\le0.4$.
 It is also observed that the results in both the two evaluation metrics first increase and then decrease along with the increase of $\delta$, where the best performance is achieved at $\delta=0.3$.
 %This is because that the margin in AUC loss is too small and thus the samples between $\mathbf{S}^{+}$ and $\mathbf{S}^{-}$ are not well differentiated.

 %\begin{table}[t]
% \centering
% \caption{Effects of hyperparameters on performance}
% \vspace{12pt}
% \label{table:delta}
% \begin{tabular}{c c c c c c}
% \toprule
% metric & 0.1 & 0.2 & 0.3 & 0.4 & 0.5  \\
% \hline
% Closed acc & 92.14\% & 0.2 & 0.3 & 0.4 & 0.5  \\
% F1 score & 0.1 & 0.2 & 0.3 & 0.4 & 0.5  \\
% \hline
% Closed acc & 92.14\% & 0.2 & 0.3 & 0.4 & 0.5  \\
% F1 score & 0.1 & 0.2 & 0.3 & 0.4 & 0.5  \\
% \bottomrule
% \end{tabular}
% \end{table}

\section{Conclusions}
\label{sec:conclusions}

 In this study, we have proposed a robust and highly accurate KWS method based on a novel multi-class AUC loss function and a confidence based decision method.
 Our KWS method not only significantly improves the robustness of the model against unseen sounds by optimizing the proposed multi-class AUC loss, but also eliminates the complex back-end processing module by using the simple confidence based decision method.
 To our knowledge, it is the first time that the low resource keyword spotting task is formulated as an open-set recognition problem.
 We compared the proposed method with four representative methods on the two public available datasets GSC v1 and GSC v2.
 Experimental results show that the proposed method significantly outperforms the four representative methods in most evaluations with smaller model sizes and less computational complexity than the latter.

% -------------------------------------------------------------------------
\bibliographystyle{IEEEbib}
\bibliography{strings,refs}

\end{document}